\documentclass[journal]{IEEEtran}
\usepackage{amsmath,amsfonts}
\usepackage{amsfonts,amssymb}
\usepackage{algorithmic}
\usepackage[ruled,linesnumbered,vlined]{algorithm2e}
\usepackage{array}
\usepackage{xcolor}
\usepackage{multirow}
\usepackage{textcomp}
\usepackage{stfloats}
\usepackage{url}
\usepackage{verbatim}
\usepackage{graphicx}
\usepackage{cite}
\usepackage{bm}
\usepackage{float}
\usepackage{subfigure}

\usepackage{booktabs}
\usepackage{ifthen}
\usepackage{tikz}
\usetikzlibrary{arrows.meta,positioning,fit,calc}

\usepackage{xcolor}
\usepackage[normalem]{ulem} 

\begin{document}

\title{Graph-Native Attention Acceleration for Attack Detection in Cyber-Physical Systems}
\author{Zhenan Feng* and Ehsan Nekouei
\thanks{Zhenan Feng and Ehsan Nekouei are with the Department of Electrical Engineering, City University of Hong Kong. E-mail: {\tt zhenafeng2@cityu.edu.hk, enekouei@cityu.edu.hk}. The work was supported by Hong Kong Research Grants Council under Project CityU 11210024. }}



\maketitle

\begin{abstract}
Cyber-physical systems (CPSs) rely on extensive networks of sensors, controllers, and actuators, making them highly vulnerable to attacks targeting measurements, control logic, and equipment operations. While graph-based attack detectors, especially graph attention models, efficiently localize such attacks by exploiting the communication and physical interaction graphs. The computational cost of edge-wise attention scales poorly with the number of interactions. This bottleneck severely limits the scalability of online monitoring in large or densely connected CPSs.
To address this challenge, we introduce GraphGHHA, a graph-native attention acceleration layer designed to reduce the inference cost of graph attention mechanisms in CPS security.  GraphGHHA consists of a local attention unit (LAU) and a global attention unit (GAU). The LAU captures the local interactions by restricting the attention computation of each node of a CPS to a limited number of its neighbors, whereas the GAU captures the system-wide interactions. Dynamic gates then weight and combine these local and global representations.
We evaluate GraphGHHA on a networked heating, ventilation, and air conditioning (HVAC) cyber-physical system under representative attack scenarios. Based on our results, the inference latency of GraphGHHA is 89.3\% smaller than that of the standard graph attention layers (for an interaction graph size of 1536), while achieving an accuracy of 0.982 in detecting attacks. These results demonstrate that the proposed attention mechanism can significantly reduce the attack detection latency without compromising accuracy.
\end{abstract}

\section{Introduction}
\subsection{Motivation}
Cyber-physical systems (CPSs) integrate sensing, communication,
computation, and control to monitor and regulate physical processes in real time \cite{cyber2}. Examples include smart factories, water distribution networks, and commercial building heating, ventilation, and air conditioning (HVAC) systems. While increased connectivity enables coordinated control and improved efficiency, it also expands the attack surface. Adversaries can compromise measurements or commands through actions such as sensor spoofing and actuator lockouts that can disrupt operations, degrade performance, or cause physical damage \cite{cyber1}.

A core challenge is that attacks propagate through system components, \emph{i.e.}, a manipulated sensor, controller, or actuator can influence the behavior of other subsystems. Conventional deep learning approaches such as multilayer perceptrons (MLPs), convolutional neural networks (CNNs), or recurrent neural networks (RNNs) do not encode the system's interaction graph in their standard forms. MLPs ignore the relational structure between nodes, CNNs assume a regular grid topology, and RNNs primarily capture sequential dependencies. As a result, these architectures have limited ability to represent the irregular communication and physical interaction topology that characterizes CPSs.

Graph-based learning is suitable for attack detection in CPSs.
Because CPS components interact through communication and physical processes, an attack on one component may affect other parts of the system. Conventional non-graph models do not represent these irregular inter-component relationships. A CPS can instead be represented as a graph in which sensors, controllers, and actuators form the nodes, while communication and physical interactions form the edges. In particular, edge-adaptive graph aggregation mechanisms can assign data-dependent weights to different neighbors, improving sensitivity to localized and stealthy attacks compared with fixed-weight graph convolutions and non-graph baselines \cite{feng2025event,IoT,graphatt}. Prior studies have shown strong detection performance (often above 95\% accuracy) for cyber-physical anomaly detection tasks, including fault detection in unmanned aerial vehicles and in-vehicle intrusion detection \cite{he2022graph,xiao2023robust}.

Despite these benefits, deploying graph-based detectors can become costly as the number of evaluated edges increases. Reducing online aggregation cost is essential for developing scalable and efficient CPS monitoring systems, particularly when a large amount of information is processed within a limited computational budget.

\subsection{Contributions}
To reduce the inference latency of graph-based attack detection in CPSs, we propose GraphGHHA, a graph aggregation mechanism designed for attack detection under strict latency constraints. Our main contributions are:

\begin{itemize}
    \item Graph-native attention acceleration: We propose GraphGHHA as a replacement for the existing graph attention layer (GAL). It consists of a local attention unit (LAU) for adjacency-constrained sparse attention and a global attention unit (GAU) for efficient system wide interaction. The LAU performs attention based on a restricted set of adjacent neighbors for each node, while the GAU performs attention using global information.

    \item Dynamic gating mechanism: We use two learnable sigmoid gates to scale the LAU and GAU outputs before concatenation, allowing their contributions to vary across nodes and input conditions.

    \item Experimental evaluation: We integrate GraphGHHA into an attack detection framework for an HVAC system. At the largest evaluated graph size, the reported inference latency is reduced by 89.3\% relative to the evaluated vanilla GAT baseline, while the F1-score changes from 0.984 to 0.982.

\end{itemize}

\subsection{Related Work}

In this subsection, we review prior studies in three areas most relevant to this paper: graph-based methods for cyber–physical security, scalable graph inference and efficient aggregation, and HVAC attack detection.

\paragraph{Graph-based Methods for Cyber-Physical Security}
Graph neural networks (GNNs) \cite{gnn} have been increasingly applied in anomaly and attack detection in cyber-physical systems because they can represent inter-component relationships. Representative studies include an event-aware graph method for industrial control processes \cite{EA-GAT2024}, a multi-head graph model for monitoring logs \cite{GAT-AD2024}, and a spatio-temporal graph model for infrastructure networks such as water distribution systems \cite{Ding2023MST}. These works demonstrate that considering interactions among system components improves detection accuracy compared with non-graph methods. However, when attention mechanisms are used in these methods to dynamically weight these interactions, the online aggregation cost can grow rapidly with neighborhood size. This can increase inference latency as the evaluated interaction set grows.

\paragraph{Scalable Graph Inference and Efficient Aggregation}
A large number of papers studied the scalability of graph learning by reducing the cost of neighborhood aggregation, \emph{e.g.}, via sampling or clustering to mitigate neighbor explosion. Representative methods include FastGCN~\cite{chen2018fastgcn}, Cluster-GCN~\cite{chiang2019cluster}, and GraphSAINT~\cite{zeng2019graphsaint}, which improve efficiency by restricting or sampling neighborhoods during training or inference. 

At the same time, recent work on efficient interaction modeling has explored hybrid designs that combine sparse computation with an efficient global mixing branch. For example, ELFATT adopts parallel heads that couple sparse computation with a global approximation to improve efficiency~\cite{elfatt}, and hybrid sparse models with token eviction have been proposed to restore direct access to selected earlier tokens while keeping overall cost low~\cite{he2025hybrid}. Dual-stage sparse schemes have also been studied to capture both local and long-range dependencies with reduced cost~\cite{DuSA}. These mechanisms are developed for sequence- or token-indexed structures and do not directly address the irregular topology and deployment requirements of cyber-physical graphs. This motivates our
graph-based aggregation design combining an LAU and a GAU through dynamic gating in CPS interaction graphs. Importantly, sampling or clustering GNN accelerations primarily target training-time scalability, while our goal is low-latency inference for edge-adaptive aggregation in CPS monitoring. Moreover, hybrid sparse-linear attention mechanisms developed for sequence data, \emph{e.g.}, text, typically define sparsity over token positions or blocks, and therefore do not directly account for graph topology. In graph-structured CPS, the acceleration mechanism should instead be designed based on the adjacency structure that encodes physical and communication interaction, rather than token order. GraphGHHA instead exploits the adjacency graph of a CPS, which captures the physical and communication interactions, to obtain sparsity patterns and limit the per-node neighbor budget to meet real-time constraints.

\paragraph{HVAC Attack Detection}
Traditional HVAC attack detection relies on rule-based thresholds and residual analysis to identify deviations from normal behavior, which provides interpretability but often triggers false alarms under varying conditions \cite{nehasil2021versatile}. Data-driven methods such as support vector machines \cite{li2019support} and random forests \cite{wang2024sensor} offer higher accuracies by learning from historical patterns, but require large labeled datasets and are sensitive to sensor noise and drift. More recent graph-based approaches use network structure to improve robustness. For instance, graph convolutional models have been used to capture dependencies among system components \cite{fan2023leveraging}. Graph attention networks (GATs) further enhance this by assigning adaptive edge weights, and combinations with recurrent models (\emph{e.g.}, GAT–LSTM) capture temporal dynamics \cite{feng2025event}. Their online computational cost motivates more efficient aggregation mechanisms for large-scale monitoring.

\section{System Model}
\subsection{Cyber-Physical System Graph Abstraction}
\begin{figure}[ht]
  \centering
  \includegraphics[width=1\linewidth]{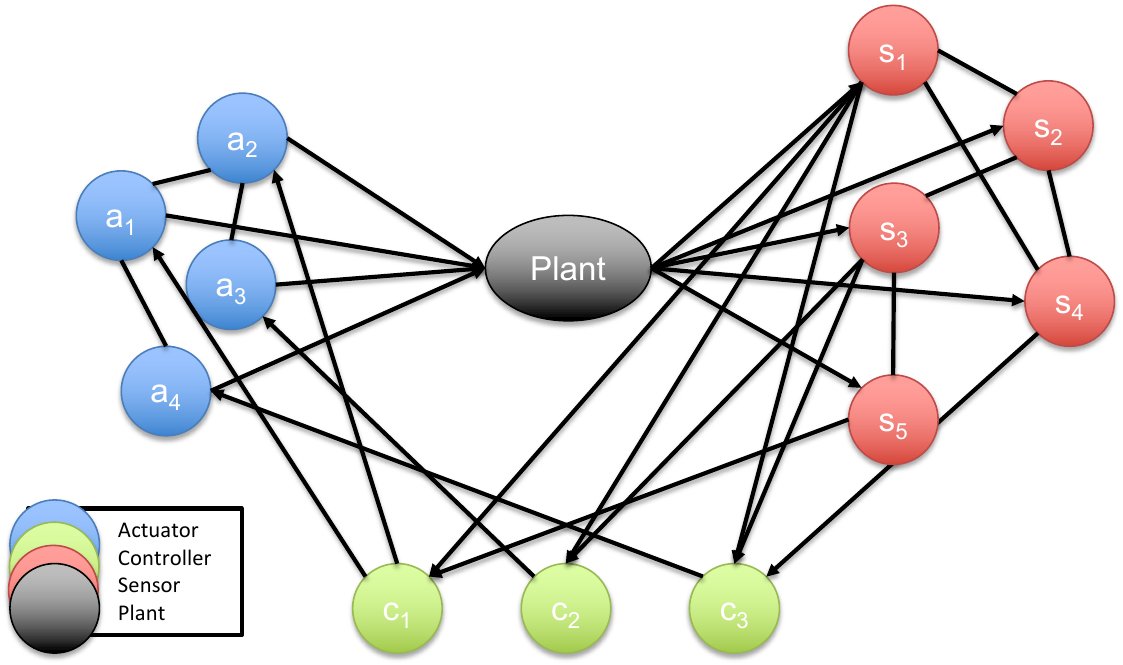}
  \caption{An example of a graph with a cyber–physical control loop. Black arrows indicate nominal measurements, control commands, and actuation signals.}
  \label{fig:auto_graph_system}
\end{figure}

A networked CPS can be abstracted as a graph $G=(\mathbb{V},\mathbb{E})$, where $\mathbb{V}$ denotes the vertex set and $\mathbb{E}$ denotes the edge set. Each node in $\mathbb{V}$ represents a functional component, \emph{e.g.}, a sensor, a controller, or an actuator. Edges in $\mathbb{E}$ encode the flow of information between components or the physical interactions among the nodes in $\mathbb{V}$. Fig.~\ref{fig:auto_graph_system} depicts the topology of a control system where sensor nodes $s_1$–$s_5$ transmit measurements over a communication network to controllers $c_1$–$c_3$, which in turn send actuation commands to actuators $a_1$–$a_4$. These actuators drive the physical plant, whose outputs are then sensed by the sensors, closing the control loop.

Formally, the vertex set of the graph is partitioned as
\begin{equation}
    \mathbb{V} = \mathbb{V}_\mathbb{S} \;\cup\; \mathbb{V}_\mathbb{C} \;\cup\; \mathbb{V}_\mathbb{A}, \nonumber
\end{equation}
where $\mathbb{V}_\mathbb{S}$, $\mathbb{V}_\mathbb{C}$, and $\mathbb{V}_\mathbb{A}$ denote sensors, controllers, and actuators, respectively.  Edges are decomposed into communication edges
\begin{equation}
    \mathbb{E}_{\mathrm{comm}} \subseteq \{(i\to j)\mid i\in \mathbb{V}_\mathbb{S}\cup \mathbb{V}_\mathbb{C},\; j\in \mathbb{V}_\mathbb{C}\cup \mathbb{V}_\mathbb{A}\},\nonumber
\end{equation}
and physical edges
\begin{equation}
    \mathbb{E}_{\mathrm{phys}} \subseteq \{(i\to j)\mid i\in \mathbb{V}_\mathbb{A},\; j\in \mathbb{V}_\mathbb{S}\}.\nonumber
\end{equation}
Thus $\mathbb{E} = \mathbb{E}_{\mathrm{comm}}\cup \mathbb{E}_{\mathrm{phys}}$ captures both logical message flows and physical interactions. 

\begin{figure*}[hb]
  \centering
  \includegraphics[width=0.7\linewidth]{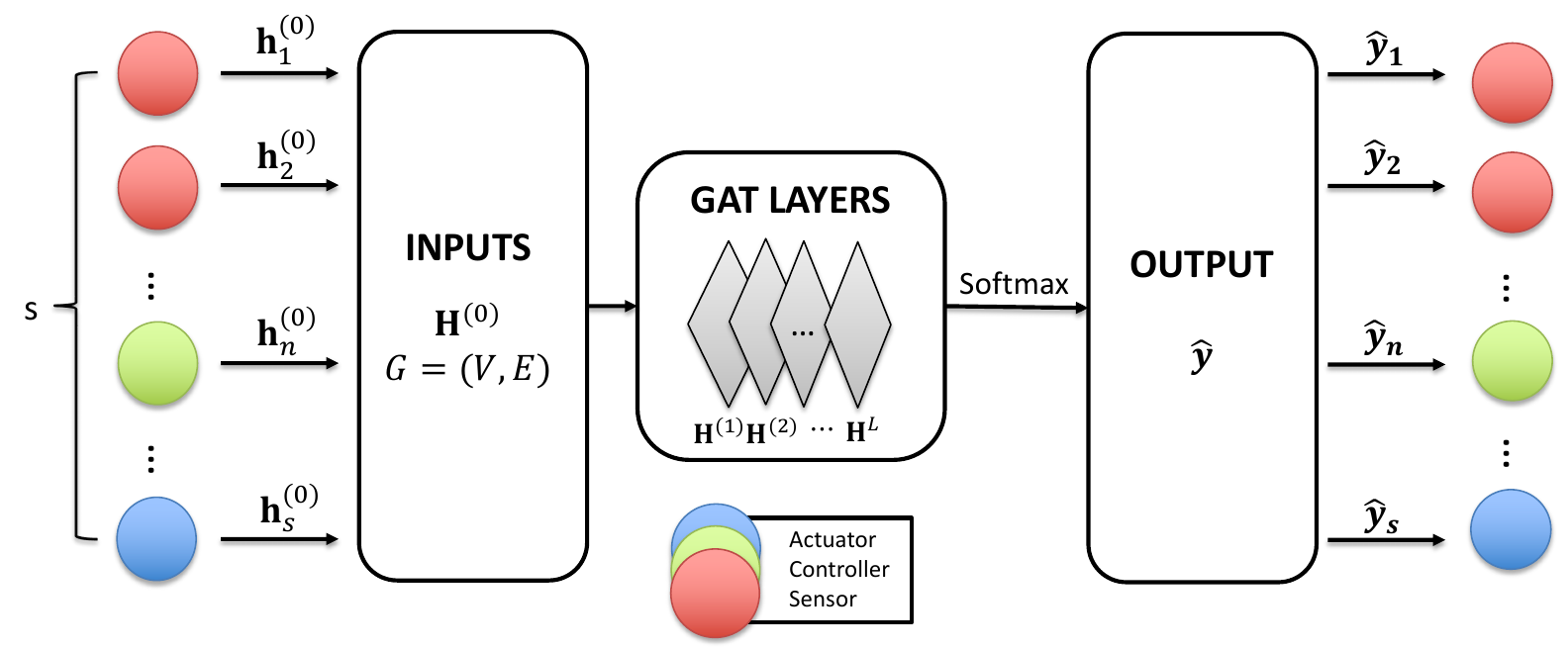}
  \caption{Attack Detection Unit (ADU).  Inputs: the initial node feature matrix $\mathbf{H}^{(0)}$ and the graph $G$.  Core: $L$ GALs.  Output: per-node attack probabilities $\hat{\mathbf{y}}$.}
  \label{fig:adu}
\end{figure*}

\subsection{Threat Model}

We assume an adversary who can attack a subset of nodes in the control system graph, \emph{e.g.}, sensors, controllers, or actuators. At the sensor level, the attacker may inject biases or drifts into measured signals. For example, an attacker may add a constant offset of $\pm2^\circ\mathrm{C}$ to a temperature measurement or inject a gradual drift into a flow measurement. Examples of the actuator-level attacks include “lockout” attack, where a valve or damper is held at a fixed position, or “leakage” attack, where false command offsets cause unintended openings of a valve or damper. Controller attacks may involve reversing control logic (\emph{e.g.}, commanding heating when cooling is required), injecting oscillatory setpoint commands, or disabling safety interlocks.

These manipulations may be sustained or temporary, and an attacker may coordinate simultaneous injections at multiple points to avoid simple residual‐based alarms. We further assume the adversary cannot modify the attack detection system, but can alter the observed feature histories of compromised nodes before it reaches the detector. Our goal is, therefore, to localize each compromised node in real-time.

\subsection{Attack Detection Unit}\label{adu}
We seek to detect whether any node $i$ is under attack in real-time. To this end, we define an Attack Detection Unit (ADU) as Fig.~\ref{fig:adu} shows. 
The initial feature vector $\mathbf{h}$ of node $i$ is: 
\begin{equation}
    \mathbf{h}_i^{(0)} \leftarrow [\,h^i_1,h^i_2,\,\dots,\,h^i_k\,]^\top \in \mathbb{R}^k.\nonumber
\end{equation}
For example, if node $i$ represents a sensor, then its feature vector $\mathbf{h}_i^{(0)}$ consists of the sensor measurements collected at that node.
Thus, the initial input feature matrix of ADU is as follows,
\begin{equation}
    \mathbf{H}^{(0)} = \bigl[\mathbf{h}_1^{(0)},\dots,\mathbf{h}_{s}^{(0)}\bigr]^\top \in \mathbb{R}^{s\times k},\nonumber
\end{equation}
where $s = |\mathbb{V}|$ denotes the total number of nodes in $\mathbb{V}$. The ADU applies $L$ graph attention layers (GAL) to refine embeddings as follows,
\begin{equation}
  \mathbf{H}^{(\ell+1)} \;=\;{\rm GAL}^{(\ell+1)}(\mathbf{H}^{(\ell)}, G).\nonumber
\end{equation}
where $\ell = [0,1,\dots,L-1]$.

In the proposed method (Section~\ref{sec:ls-gat}), we keep the ADU architecture unchanged but replace the ${\rm GAL}$ block with the proposed accelerated graph layer GraphGHHA to enable fast inference on large connected control graphs.

Then, the ADU outputs a vector of per-node attack probabilities
\begin{equation}
  \hat{\mathbf{y}} \;=\;\bigl[\hat y_1,\dots,\hat y_{s}\bigr]^\top
  \;\in\;[0,1]^{s}\,,\nonumber
\end{equation}
where $\hat y_i$ denotes the predicted probability that node $i$ is under attack. More specifically, for binary detection at each node, the ADU applies a linear classifier followed by a sigmoid function:
\begin{equation}
\hat{\mathbf y}=\operatorname{Sigmoid}
 (\mathbf H^{(L)}\mathbf w_o+b_o\mathbf 1_s),
\qquad \mathbf w_o\in\mathbb R^k.
\nonumber
\end{equation}
Here $\mathbf w_o$ and $b_o$ are learnable classifier parameters.

In addition to detection, we also assign labels for training that indicate the specific type of injected attack, \emph{e.g.}, $j \in \{0,1,2,3\}$, where $0=\text{normal},\; 1=\text{actuator attack},\; 2=\text{sensor attack},\; 3=\text{controller attack}$.
This allows the ADU to be extended to a multi-class classification task. In this case, the predicted label of node $i$ is obtained by
\begin{equation}
   \hat{c}_i \;=\;\arg\max_j \hat{y}_{i,j}, \nonumber
\end{equation}
where $\hat{y}_{i,j}$ denotes the probability of node $i$ being in attack class $j$. Although classification of attack types is not the main focus of this study, this formulation shows the ADU to support both binary detection and multi-class identification.

\subsection{Graph Attention Layer}
Each ${\rm GAL}^{(\ell+1)}$ ($\ell= [0,1,\dots,L-1]$) updates the node embeddings by computing edge‐wise attention and aggregating neighbor features:
\begin{align}
  e_{ij}^{(\ell+1)}
  &= \mathrm{LeakyReLU}\Bigl(a^{(\ell+1)\top}\bigl[\mathcal{W}_{\rm GAL}^{(\ell+1)}\mathbf{h}_i^{(\ell)}\;\|\;\mathcal{W}_{\rm GAL}^{(\ell+1)}\mathbf{h}_j^{(\ell)}\bigr]\Bigr), \label{eq1}\\
  \mathcal{A}_{ij}^{(\ell+1)}
  &= \frac{\exp\bigl(e_{ij}^{(\ell+1)}\bigr)}
          {\sum_{k\in\mathcal{N}(i)}\exp\bigl(e_{ik}^{(\ell+1)}\bigr)}, \label{eq2}\\
  \mathbf{h}_i^{(\ell+1)}
  &= \sigma\Bigl(\sum_{j\in\mathcal{N}(i)}\mathcal{A}_{ij}^{(\ell+1)}\,\mathcal{W}_{\rm GAL}^{(\ell+1)}\mathbf{h}_j^{(\ell)}\Bigr),\label{eq3}
\end{align}
where $\mathcal{N}(i)$ is the neighborhood of node $i$, $\mathcal{W}_{\rm GAL}^{(\ell)}\in\mathbb{R}^{k\times k}$ and ${a^{(\ell)}\in\mathbb{R}^{2k}}$ are learnable, and $\sigma(\cdot)$ is a nonlinear activation function.
Let $\mathfrak{A}\in\{0,1\}^{s\times s}$ denote the (unweighted) adjacency matrix of the graph $G$, where $\mathfrak{A}_{ij}=1$ iff $(i\to j)\in\mathbb{E}$. Then $\mathcal{N}(i)=\{j\mid \mathfrak{A}_{ij}=1\}$, \emph{i.e.}, attention is only evaluated on graph edges.
The corresponding matrix form of equation~\eqref{eq3} is as follows,
\begin{equation}
    \quad \mathbf{H}^{(\ell+1)}\leftarrow\sigma\bigl(\mathcal{A}^{(\ell+1)}\mathbf{H}^{(\ell)}{\mathcal{W}_{\rm GAL}^{(\ell+1)\top}\bigr)}.
    \label{eq4}
\end{equation}
where $\mathcal{A}$ denotes the weighted adjacency matrix by performing attention.
In practice, $\mathcal{A}^{(\ell+1)}$ inherits the sparsity pattern of $\mathfrak{A}$, \emph{i.e.}, $\mathcal{A}^{(\ell+1)}_{ij}=0$ when $\mathfrak{A}_{ij}=0$.

\providecommand{\GHdel}[1]{{\color{blue}\sout{#1}}}
\providecommand{\GHadd}[1]{{\color{red}#1}}
\providecommand{\GHdelmath}[1]{{\color{blue}\sout{\mbox{$\displaystyle #1$}}}}
\newsavebox{\GHreviewbox}
\newcommand{\GHdelrow}[1]{%
  \par\begingroup
  \sbox{\GHreviewbox}{\GHdelmath{#1}}%
  \noindent\ifdim\wd\GHreviewbox>\linewidth
    \resizebox{\linewidth}{!}{\usebox{\GHreviewbox}}%
  \else
    \makebox[\linewidth][c]{\usebox{\GHreviewbox}}%
  \fi\par\endgroup}

\section{GraphGHHA: Graph-Native Hybrid-Head Acceleration for GAL}\label{sec:ls-gat}

As discussed above, the ADU adopts GALs to capture component relationships in CPSs. However, online edge-adaptive aggregation becomes expensive when the interaction graph is large or densely connected. In general, the aggregation cost scales with the number of evaluated edges as $O(|\mathbb E|k)$, where $|\mathbb E|$ is the number of edges and $k$ is the feature dimension. For dense graphs with $|\mathbb E|=O(s^2)$ ($s=|\mathbb V|$), the cost becomes quadratic in $s$, leading to high inference latency that can delay real-time alarms.

To address this, we propose GraphGHHA, a graph-native attention acceleration to replace GAL. GraphGHHA combines two parallel units: the LAU performs restricted graph-local attention, and the GAU aggregates system-wide information. Two learnable sigmoid gates scale the unit outputs before concatenation, preserving accuracy while reducing inference time.

In this section, we first summarize scaled dot-product attention and associative global mixing. We then describe the LAU, GAU, and dynamic gating mechanism of GraphGHHA, followed by the forward-pass algorithm. Finally, we analyze computational cost and mixed-precision inference.

\subsection{Preliminaries}
\subsubsection{Vanilla Scaled Dot-Product Attention}
Let $\mathbf{H}^{(\ell)}\in\mathbb{R}^{s\times k}$ be the input feature matrix for $s$ nodes of dimensionality $k$. The vanilla scaled dot‐product attention (VSDPA) computes
\begin{align}
\mathbf{Q}^{(\ell+1)} &= \mathbf{H}^{(\ell)}\mathcal{W}^{(\ell+1)}_Q, \label{eq:proj1}\\
\mathbf{K}^{(\ell+1)} &= \mathbf{H}^{(\ell)}\mathcal{W}^{(\ell+1)}_K, \label{eq:proj2}\\
\mathbf{V}^{(\ell+1)} &= \mathbf{H}^{(\ell)}\mathcal{W}^{(\ell+1)}_V, \label{eq:proj3}\\
\mathbf{A}^{(\ell+1)} &= \mathrm{Softmax}\bigl(\mathbf{Q}^{(\ell+1)}\,\mathbf{K}^{(\ell+1)\top} / \sqrt{k}\bigr),
\label{eq:attn}\\
\mathbf{H}^{(\ell+1)} &= \mathbf{A}^{(\ell+1)}\,\mathbf{V}^{(\ell+1)},
\label{eq:agg}
\end{align}
where $\mathcal{W}_Q\in\mathbb{R}^{k\times k}$, $\mathcal{W}_K\in\mathbb{R}^{k\times k}$, and $\mathcal{W}_V\in\mathbb{R}^{k\times k}$ are three scaling parameter matrices. VSDPA forms the attention matrix by normalizing pairwise similarities and aggregates values as in \eqref{eq:proj1}-\eqref{eq:agg}. Its dominant cost comes from forming an $s\times s$ similarity matrix, leading to quadratic complexity in the number of elements $s$ \cite{CURSA,DuSA}. This motivates linearization and sparsification methods used by GraphGHHA.

\subsubsection{Kernelized Reformulation}
A common approach to avoid explicit $s\times s$ softmax attention is to use a nonnegative feature map $\phi(\cdot)$ to obtain a linear attention form computed via associative products, \emph{e.g.}, $\phi(\mathbf{Q})\big(\phi(\mathbf{K})^\top\mathbf{V}\big)$ \cite{transformersarernn}. 
GraphGHHA adopts this idea in its GAU, while preserving graph-masked softmax attention in its LAU.

\subsubsection{Multi-Head Design}
Multi-head attention improves expressivity by using $p$ parallel heads and concatenating their outputs, as in \eqref{eq:multiattn} and \eqref{eq:multiagg}. Each GraphGHHA head contains an LAU and a GAU to balance efficiency and accuracy.

\begin{align}
\mathbf{A}_{i}^{(\ell+1)} &= \mathrm{Softmax}\bigl(\mathbf{Q}_{i}^{(\ell+1)}\,\mathbf{K}_{i}^{(\ell+1)\top} / \sqrt{k/p}\bigr),
\label{eq:multiattn}\\
\mathbf{H}^{(\ell+1)} &= \bigl[\mathbf{A}_{1}^{(\ell+1)}\,\mathbf{V}_{1}^{(\ell+1)}, \mathbf{A}_{2}^{(\ell+1)}\,\mathbf{V}_{2}^{(\ell+1)},\dots, \mathbf{A}_{p}^{(\ell+1)}\,\mathbf{V}_{p}^{(\ell+1)}\bigr],
\label{eq:multiagg}
\end{align}
where \mbox{$\mathbf{Q}_{i}^{(\ell+1)}\in\mathbb{R}^{s \times (k/p)}$}, \mbox{$\mathbf{K}_{i}^{(\ell+1)}\in\mathbb{R}^{s \times (k/p)}$}, \mbox{$i=[1,2,\dots,p]$}, and \mbox{$p$} is the number of heads. In what follows, we describe one representative head and omit its head index. $i,j$ are used to denote graph nodes.

\subsection{The Proposed GraphGHHA Layer}
\label{GraphGHHA}

Each head in GraphGHHA contains a GAU and an LAU arranged in parallel. The GAU aggregates information across all nodes through associative matrix products, while the LAU performs softmax attention over retained graph neighbors. These units constitute the global and local branches of the head, respectively.

We describe one representative head below and omit its head index for simplicity. The layer input is $\mathbf H=\mathbf H^{(\ell)}\in\mathbb R^{s\times k}$, where $k$ is the total feature width. With $p$ heads, each head has width $k/p$ and allocates $k_1$ channels to the GAU and $k_2$ channels to the LAU, where $k_1+k_2=k/p$ and $k_1,k_2\ge1$. Subscripts 1 and 2 denote the GAU and LAU, respectively. For simplicity, we omit the layer index $\ell$ when describing
one head and include it again when writing the layer output

\paragraph{Feature projection and channel split}
The query, key, and value matrices are obtained from the layer input:
\begin{equation}
\mathbf Q=\mathbf H\mathcal W_Q,\qquad
\mathbf K=\mathbf H\mathcal W_K,\qquad
\mathbf V=\mathbf H\mathcal W_V,
\label{eq:graphghha_projections}
\end{equation}
where $\mathcal W_Q,\mathcal W_K,\mathcal W_V \in\mathbb R^{k\times(k/p)}$ are this head's learnable projection matrices. They can equivalently be implemented as column blocks of the full $k\times k$ layer projections.
We split $\mathbf Q=[\mathbf Q_1,\mathbf Q_2]$, and similarly for $\mathbf K$ and $\mathbf V$, with $\mathbf Q_b,\mathbf K_b,\mathbf V_b\in\mathbb R^{s\times k_b}$ for $b\in\{1,2\}$ and $k_1+k_2=k/p$.

\paragraph{Global attention unit (GAU)}
The GAU computes
\begin{align}
\mathbf M_1&=\phi(\mathbf K_1)^\top\mathbf V_1,
\label{eq:gmu_summary}\\
\mathbf H_1&=\phi(\mathbf Q_1)\mathbf M_1,
\qquad\phi(x)=(\max(0,x))^2.
\label{eq:gmu_output}
\end{align}
The feature map $\phi$ is applied element-wise. Computing $\mathbf M_1\in\mathbb R^{k_1\times k_1}$ first avoids forming an $s\times s$ interaction matrix and produces $\mathbf H_1\in\mathbb R^{s\times k_1}$. This operation provides unnormalized global mixing.

\paragraph{Local attention unit (LAU)}
The LAU uses the CPS graph to restrict attention to relevant physical and communication relationships. Let $\mathcal N(i)$ denote the neighborhood of node $i$ defined in the system model. The neighbor-budget parameter $n\in\{1,\ldots,s\}$ sets the maximum retained neighborhood size to $r=\lfloor s/n\rfloor$. Following a fixed node-index order, we keep the first $\min\{r,|\mathcal N(i)|\}$ neighbors to form $\mathcal N_r(i)$. The retained sets are shared across heads. 

For each retained neighbor, the LAU computes
\begin{align}
    e_{ij}^{\mathrm{loc}}
    &=\frac{\mathbf q_{2,i}^{\top}\mathbf k_{2,j}}{\sqrt{k_2}},
    \qquad j\in\mathcal N_r(i),
    \label{eq:lau_score}\\
    \alpha_{ij}^{\mathrm{loc}}
    &=\frac{\exp(e_{ij}^{\mathrm{loc}})}
    {\sum_{u\in\mathcal N_r(i)}\exp(e_{iu}^{\mathrm{loc}})},
    \label{eq:lau_weight}\\
    \mathbf h_{2,i}
    &=\sum_{j\in\mathcal N_r(i)}
    \alpha_{ij}^{\mathrm{loc}}\mathbf v_{2,j}.
\label{eq:fastattngraph}
\end{align}
Here, $\mathbf q_{2,i}$, $\mathbf k_{2,j}$, and $\mathbf v_{2,j}\in\mathbb R^{k_2}$ are the corresponding row features written as column vectors. The vector $\mathbf h_{2,i}^{\top}$ forms row $i$ of $\mathbf H_2\in\mathbb R^{s\times k_2}$. 

\paragraph{Dynamic gating and fusion}
The two gates are computed from the layer input:
\begin{equation}
    [\mathbf C_1,\mathbf C_2] =\operatorname{Sigmoid}(\mathbf H\mathcal W_C),
\label{eq:graphghha_gates}
\end{equation}
where $\mathcal W_C\in\mathbb R^{k\times2}$ and $\mathbf C_1,\mathbf C_2\in\mathbb R^{s\times1}$. The same two gates are shared across heads. The GAU and LAU outputs within the representative head are combined as
\begin{equation}
\begin{aligned}
    \mathbf Z=\bigl[
    &\operatorname{Broad}_{k_1}(\mathbf C_1)\odot\mathbf H_1,\\
    &\operatorname{Broad}_{k_2}(\mathbf C_2)\odot\mathbf H_2\bigr]\in\mathbb R^{s\times(k/p)}.
\end{aligned}
\label{eq:fastattnsingle}
\end{equation}
Here, $\operatorname{Broad}_{q}$ broadcasts a gate across $q$ feature channels, $\odot$ denotes element-wise multiplication, and brackets denote concatenation along the feature dimension. The two gates independently scale the GAU and LAU outputs.

\paragraph{Layer output}
As in the multi-head design, the outputs of the $p$ hybrid heads are concatenated along the feature dimension to form $\mathbf H_{\mathrm{cat}}^{(\ell+1)}\in\mathbb R^{s\times k}$. Restoring the layer indices, the output is obtained as follows:
\begin{align}
    \mathbf H_{\mathrm{att}}^{(\ell+1)} &=\mathbf H_{\mathrm{cat}}^{(\ell+1)}\mathcal W_H^{(\ell+1)} +\mathbf H^{(\ell)}, \label{eq:graphghha_residual}\\
    \mathbf H^{(\ell+1)} &=\operatorname{MLP}\bigl(\operatorname{LN} (\mathbf H_{\mathrm{att}}^{(\ell+1)})\bigr) +\mathbf H_{\mathrm{att}}^{(\ell+1)}. \label{eq:graphghha_output}
\end{align}
Here, $\mathcal W_H^{(\ell+1)}\in\mathbb R^{k\times k}$ is a learnable matrix, $\operatorname{LN}$ denotes layer normalization, and $\operatorname{MLP}$ denotes a multilayer perceptron applied independently to each node with output width $k$.
The output $\mathbf H^{(\ell+1)}\in\mathbb R^{s\times k}$ combines local and global information before being passed to the next block.

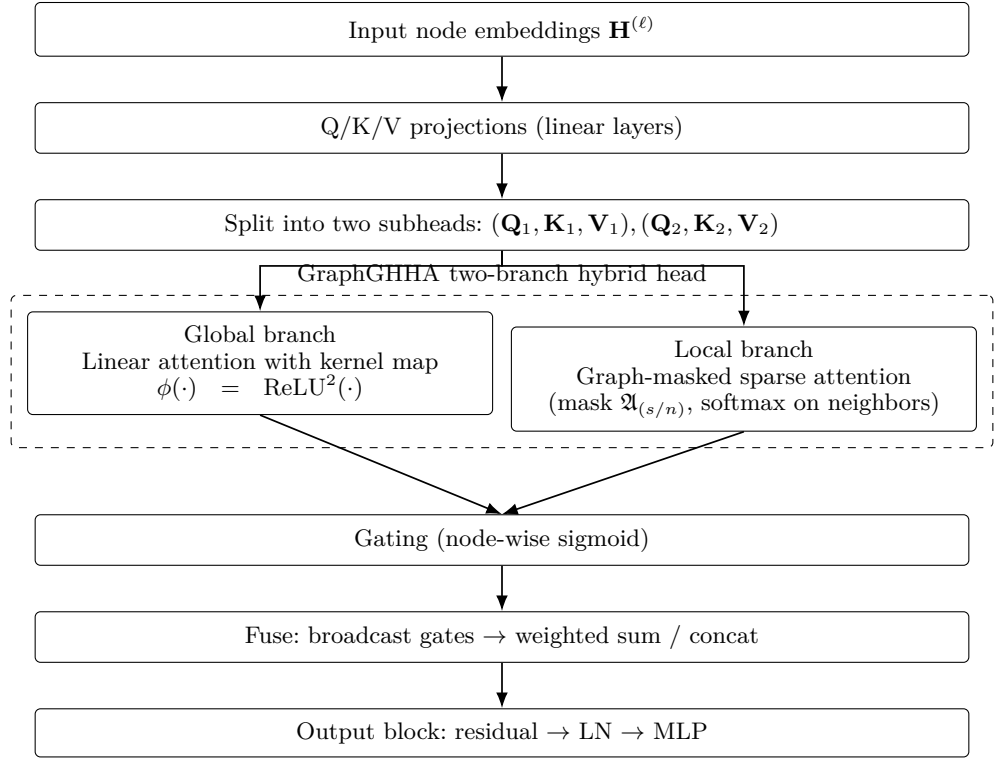
\begin{figure*}[!t]
\centering
\begin{minipage}{0.78\textwidth}
\centering
\begin{tikzpicture}[
    font=\small,
    node distance=6mm and 10mm,
    arr/.style={-Latex, line width=0.65pt},
    box/.style={draw, rounded corners=2pt, align=center, inner sep=5pt, text width=12.0cm},
    op/.style={draw, rounded corners=2pt, align=center, inner sep=5pt, text width=12.0cm},
    twobox/.style={draw, rounded corners=2pt, align=center, inner sep=5pt, text width=5.8cm}
]

\node[box] (H) {Input node embeddings $\mathbf{H}^{(\ell)}$};

\node[op, below=of H] (proj)
{Q/K/V projections (linear layers)};

\node[op, below=of proj] (split)
{{One head: allocate channels to the GAU and LAU\\ $k_1+k_2=k/p$}};

\draw[arr] (H) -- (proj);
\draw[arr] (proj) -- (split);

\node[twobox, below=8mm of split, xshift=-3.2cm] (global)
{\textbf{Global attention unit (GAU)}\\
Associative global mixing with kernel map\\
$\phi(\cdot)=\mathrm{ReLU}^2(\cdot)$};

\node[twobox, below=10mm of split, xshift= 3.2cm] (local)
{\textbf{Local attention unit (LAU)}\\
Graph-masked sparse attention\\
{(softmax over $\mathcal N_r(i)$, $r=\lfloor s/n\rfloor$)}};

\draw[arr] (split.south) -- ++(0,-2mm) -| (global.north);
\draw[arr] (split.south) -- ++(0,-2mm) -| (local.north);

\node[op, below=12mm of $(global.south)!0.5!(local.south)$] (gate)
{{Two dynamic gates computed from $\mathbf H$\\}};

\draw[arr] (global.south) -- (gate.north);
\draw[arr] (local.south)  -- (gate.north);

\node[op, below=of gate] (fuse)
{{Fuse: gate and concatenate the GAU and LAU outputs to form $\mathbf Z$}};

\node[box, below=of fuse] (out)
{{Concatenate head outputs: residual $\rightarrow$ LN $\rightarrow$ MLP}};

\draw[arr] (gate) -- (fuse);
\draw[arr] (fuse) -- (out);

\node[draw, dashed, rounded corners=2pt, fit=(global)(local), inner sep=6pt,
      label={[font=\small]above:}] {};

\end{tikzpicture}
\end{minipage}
\caption{Simplified workflow of one GraphGHHA layer (equations are provided in the main text).}
\label{fig:graphghha_workflow_simple}
\end{figure*}

\begin{algorithm}[ht]
\caption{GraphGHHA Forward Pass}
\label{alg:mha_block}
\KwIn{Node embeddings $\mathbf H^{(\ell)}\in\mathbb R^{s\times k}$; $p$ heads with $k_1+k_2=k/p$; neighbor lists $\{\mathcal N_r(i)\}_{i=1}^s$}
\KwOut{Updated embeddings $\mathbf H^{(\ell+1)}\in\mathbb R^{s\times k}$}
\BlankLine
$\mathbf H\leftarrow\mathbf H^{(\ell)}$\;
$[\mathbf C_1,\mathbf C_2]\leftarrow
\operatorname{Sigmoid}(\mathbf H\mathcal W_C)$\;
\tcp{The following operations describe one representative head}
$\mathbf Q\leftarrow\mathbf H\mathcal W_Q$,
$\mathbf K\leftarrow\mathbf H\mathcal W_K$,
$\mathbf V\leftarrow\mathbf H\mathcal W_V$\;
Split $\mathbf Q,\mathbf K,\mathbf V$ into GAU channels
(width $k_1$) and LAU channels (width $k_2$)\;
\tcp{Global attention unit}
$\mathbf M_1\leftarrow\phi(\mathbf K_1)^\top\mathbf V_1$\;
$\mathbf H_1\leftarrow\phi(\mathbf Q_1)\mathbf M_1$\;
\tcp{Local attention unit}
$\mathbf H_2\leftarrow\mathbf0_{s\times k_2}$\;
\For{$i\leftarrow1$ \KwTo $s$}{
 \If{$\mathcal N_r(i)\ne\varnothing$}{
  $e_{ij}^{\mathrm{loc}}\leftarrow
  \mathbf q_{2,i}^{\top}\mathbf k_{2,j}/\sqrt{k_2}$
  for $j\in\mathcal N_r(i)$\;
  $\alpha_{ij}^{\mathrm{loc}}\leftarrow
  \exp(e_{ij}^{\mathrm{loc}})/
  \sum_{u\in\mathcal N_r(i)}\exp(e_{iu}^{\mathrm{loc}})$
  for $j\in\mathcal N_r(i)$\;
  $\mathbf H_2[i,:]\leftarrow
  \sum_{j\in\mathcal N_r(i)}
  \alpha_{ij}^{\mathrm{loc}}\mathbf V_2[j,:]$\;
 }
}
\tcp{Gated output of this head}
$\mathbf Z\leftarrow
[\operatorname{Broad}_{k_1}(\mathbf C_1)\odot\mathbf H_1,
\operatorname{Broad}_{k_2}(\mathbf C_2)\odot\mathbf H_2]$\;
Apply the same operations to all $p$ heads using their respective
projection matrices, and concatenate their outputs into
$\mathbf H_{\mathrm{cat}}^{(\ell+1)}\in\mathbb R^{s\times k}$\;
$\mathbf H_{\mathrm{att}}^{(\ell+1)}\leftarrow
\mathbf H_{\mathrm{cat}}^{(\ell+1)}\mathcal W_H^{(\ell+1)}
+\mathbf H^{(\ell)}$\;
$\mathbf H^{(\ell+1)}\leftarrow
\operatorname{MLP}\bigl(\operatorname{LN}
(\mathbf H_{\mathrm{att}}^{(\ell+1)})\bigr)+\mathbf H_{\mathrm{att}}^{(\ell+1)}$\;
\Return{$\mathbf H^{(\ell+1)}$}
\end{algorithm}

Algorithm~\ref{alg:mha_block} details the operations. The GAU computes a global summary, while the LAU aggregates only retained graph neighbors using sparse neighbor lists. The gates scale the GAU and LAU outputs. The head outputs are then concatenated and passed through the output projection and residual updates.

\subsection{Complexity Analysis}
\label{ComANA}
We first analyze one GraphGHHA head operating on $s$ nodes, with GAU and LAU widths $k_1$ and $k_2$, where $k_1+k_2=k/p$. The maximum retained neighborhood size is $r=\lfloor s/n\rfloor$. The aggregation cost is divided into two parts:
\begin{equation}
    \mathrm{Cost}_{\mathrm{head}}
    =C_{\mathrm{global}}+C_{\mathrm{local}}.
\end{equation}
Here, $C_{\mathrm{global}}$ and $C_{\mathrm{local}}$ denote the costs of the GAU and LAU, respectively, including output scaling by their gates.
\paragraph{Global Attention unit (GAU)}
The gated GAU output is
\begin{equation}
\begin{aligned}
    \operatorname{Broad}_{k_1}(\mathbf C_1)\odot
    \bigl[\phi(\mathbf Q_1)
    \bigl(\phi(\mathbf K_1)^\top\mathbf V_1\bigr)\bigr].
\end{aligned}
\end{equation}
Applying the squared ReLU feature map and scaling the output by the broadcast gate each cost $O(sk_1)$. Computing $\phi(\mathbf K_1)^\top\mathbf V_1$ first and then multiplying by $\phi(\mathbf Q_1)$ costs $O(sk_1^2)$.
Hence,
\begin{equation}
    C_{\mathrm{global}}
    =O(sk_1)+O(sk_1^2)+O(sk_1)
    =O(sk_1^2).
\end{equation}
For fixed $k_1$, the GAU has linear complexity in $s$.

\paragraph{Local attention unit (LAU)}
The gated LAU output is
\begin{equation}
    \operatorname{Broad}_{k_2}(\mathbf C_2)\odot\mathbf H_2,
\end{equation}
where $\mathbf H_2$ is computed by \eqref{eq:lau_score}-\eqref{eq:fastattngraph}.
Evaluating scores over at most $r$ retained neighbors per node costs $O(srk_2)$, and neighbor-wise normalization costs $O(sr)$. Aggregating the corresponding value vectors costs $O(srk_2)$. Initializing the local output and applying the broadcast gate cost $O(sk_2)$. Thus, the worst-case cost of the LAU is
\begin{equation}
\begin{aligned}
    C_{\mathrm{local}}
    &=O(srk_2)+O(sr)+O(srk_2)+O(sk_2)\\
    &=O(srk_2)=O\!\left(\frac{s^2}{n}k_2\right).
\end{aligned}
\end{equation}
If $s/n\ll s$ and $k_2<s$, this reduces to $C_{\mathrm{local}}\approx O(s)$.

\paragraph{Overall scaling}
Summing the costs of all $p$ heads gives
\begin{equation}
\begin{aligned}
    \mathrm{Cost}_{\mathrm{aggregation}}
    &=p\bigl(C_{\mathrm{global}}+C_{\mathrm{local}}\bigr)\\
    &=O(psk_1^2+psrk_2).
\end{aligned}
\end{equation}
Including the dense input/output projections and an MLP with a fixed expansion ratio, the per-layer cost is
\begin{equation}
    \mathrm{Cost}_{\mathrm{layer}}
    =O(sk^2+psk_1^2+psrk_2).
\label{eq:graphghha_cost}
\end{equation}

Following the above near-linear GAU and LAU cost estimates, for fixed $p$ and feature widths,
\begin{equation}
    \mathrm{Cost}_{\mathrm{layer}}
    \approx O(s)+p\bigl[O(s)+O(s)\bigr]=O(s).
\end{equation}
By choosing $k_1,k_2<s$ and $s/n\ll s$, GraphGHHA achieves near-linear complexity in $s$, versus the $O(s^2)$ aggregation cost of dense all-pairs softmax attention at fixed feature width. Moreover, when the ratio $s/n$ decreases and $s\gg k_1,k_2$, the acceleration effect becomes increasingly significant.

\subsection{Quantization in GraphGHHA}
To further accelerate inference and reduce memory bandwidth, we adopt mixed-precision arithmetic, storing all model parameters and intermediate activations in reduced-precision floating-point format (\emph{e.g.},\ FP16 or BF16).  Each projection weight matrix $\mathcal{W}$ is cast from full-precision FP32 to low-precision $\mathrm{FP}_e$ (FP16 or BF16) via a single hardware-accelerated conversion:
\begin{equation}
    \mathcal{W}_{e} = \mathrm{cast}_{\mathrm{FP}_e}(\mathcal{W}),\nonumber
\end{equation}
and the feature matrices $\mathbf X\in\{\mathbf Q,\mathbf K,\mathbf V\}$ are similarly converted:
\begin{equation}
    \mathbf{X}_{e} = \mathrm{cast}_{\mathrm{FP}_e}(\mathbf{X}).\nonumber
\end{equation}
All matrix multiplications and additions are executed in $\mathrm{FP}_e$ and other operations such as the Softmax operation and loss computation are still executed in FP32. When combined with structured sparsification, FP16 mixed precision reduces memory traffic by half and speeds inference by about 1.5 times on modern GPUs while incurring only a 0.2 percentage point drop in F1-score from 0.978 to 0.976 in our HVAC case study (see Section~\ref{quantization}).

\section{Simulation Results}
\subsection{Experimental Setup}
We evaluate GraphGHHA using an HVAC CPS modeled in HVACSIM$+$. Each zone $i$ is served by its own FCU comprising temperature sensors, flow sensors, valve and damper actuators, and a central controller that issues setpoints to maintain thermal comfort. The system is represented as a graph $G=(\mathbb{V},\mathbb{E})$, where each zone node is connected to its associated sensors, actuators, and controller. The base graph contains $s=32$ nodes.
To evaluate scalability, we consider enlarged graph instances with sizes ranging from 32 to 1536 nodes. Features from different component types are projected into a shared latent space, and the neighbor-budget parameter is fixed at $n=8$. We evaluate detection performance and inference latency across these graph sizes under attacks affecting sensors, actuators, and controllers.

\subsubsection*{Dataset}
We evaluate the proposed method using datasets from \cite{lbnl}. Each zone is controlled to maintain a setpoint of 22°C during occupied hours and 28°C during setbacks. Data from 29 different signal types (temperature, flow, valve and fan positions, etc.) are logged at 1 minute intervals for one year (525600 samples), recording sensor readings, control commands, and actuator positions. The control loop consists of: 
\begin{itemize}
  \item \textbf{Sensors:} room air temperature sensor, supply water temperature sensor, supply air flow sensor, zone humidity sensor.
  \item \textbf{Controllers:} fan controller, outdoor air damper controller, cooling coil valve controller, heating coil valve control sequence, and zone temperature setpoints were set by PID control sequences.
  \item \textbf{Actuators:} modulating water valves on the cooling and heating coils, a variable‐speed fan, and an outdoor‐air damper.
\end{itemize}

\begin{table*}[ht]
  \centering
  \caption{Latency (ms) and reduction rate (\%) vs.\ node size $s'$, with reduction measured relative to Vanilla GAT.}
  \label{tab:throughput_latency}
  \small
  \begin{tabular}{r||
                  c||
                  c|c||
                  c|c||
                  c|c}
    \toprule
    \multirow{2}{*}{$s'$} 
      & \multicolumn{1}{c||}{\textbf{Vanilla GAT}} 
      & \multicolumn{2}{c||}{\textbf{Sparsified GAT}} 
      & \multicolumn{2}{c||}{\textbf{Quantized GAT}} 
      & \multicolumn{2}{c}{\textbf{GraphGHHA}} \\
    & ms  & ms & Reduction (\%) & ms & Reduction (\%) & ms & Reduction (\%) \\
    \midrule
     32   & 0.0113    & 0.0164 & -45.13 & 0.01381 & -22.21 & 0.0108 & 4.42 \\
     64   & 0.0147    & 0.0197 & -34.01 & 0.01398 & 4.90   & 0.0139 & 5.44 \\
    128   & 0.0254    & 0.0328 & -29.13 & 0.01537 & 39.49  & 0.0178 & 29.92 \\
    256   & 0.0636    & 0.0596 & 6.29   & 0.0315  & 50.47  & 0.0292 & 54.09 \\
    512   & 0.207     & 0.1100 & 46.86  & 0.0764  & 63.09  & 0.0502 & 75.75 \\
   1024   & 0.726     & 0.2106 & 70.99  & 0.2530  & 65.15  & 0.0970 & 86.64 \\
   1536   & 1.370     & 0.3160 & 76.93  & 0.9590  & 30.00  & 0.1460 & 89.34 \\
    \bottomrule
  \end{tabular}
\end{table*}

\subsubsection*{Cyber-Physical Attack Scenarios}
We simulate 49 cyber-physical attack cases grouped into six scenario families: (i) damper attacks, including forced leakage and stuck positions of 20\%-80\% open; (ii) valve actuator attacks, including leakage offsets of 20\%--80\% and stuck commands of 0\%, 50\%, or 100\%; (iii) sensor-bias attacks, with offsets of $\pm2^\circ\mathrm{C}$ or $\pm4^\circ\mathrm{C}$ injected into zone-temperature measurements; (iv) coil-fouling attacks, modeled by reducing heat-transfer coefficients on the air or water side; (v) airflow-obstruction attacks, including filter restriction, outdoor-air inlet blockage, and fan outlet blockage; and (vi) controller attacks, including reverse-acting control signals and unstable FCU control logic.

Sensor-bias and controller attacks represent cyber-level manipulation of measurements and control logic, while coil fouling and airflow obstruction represent deliberate physical tampering with HVAC equipment. Damper and valve cases represent malicious actuator manipulation through leakage or forced operating positions. Each case is applied independently, and the model performs detection of the resulting abnormal operating state. All experiments are conducted on an NVIDIA A100 GPU.

\subsubsection*{Evaluation Metrics}  
To assess the performance, we report:
\begin{itemize}
  \item \textbf{Latency}: the average time (in milliseconds) required to detect the attack
  \item \textbf{Detection performance:} the standard classification metrics including precision, recall, and F1‐score which are computed per attack type and averaged macro‐wise.
\end{itemize}
This setup allows us to compare vanilla GAT and GraphGHHA under simulated building operating conditions and adverse measurement scenarios.  

\subsection{Latency Performance}

We compare inference latency of the following four attack detection methods:
\begin{itemize}
  \item \textbf{Vanilla GAT}\cite{graphatt}
  \item \textbf{Sparsified GAT} which only performs vanilla attention with a limited neighbor size.
  \item \textbf{Quantized GAT} which performs vanilla GAT implemented in Quantized floating-point arithmetic.
  \item \textbf{GraphGHHA} which performs the proposed Graph Gated Hybrid Head Attention Mechanism.
\end{itemize}

All four methods use three stacked attention layers followed by a linear classification head. We measure inference latency as a function of the graph node count $s^{\prime}$.

Table~\ref{tab:throughput_latency} shows the average latency (ms) and the latency reduction rate (\%) for four attention variants as the node size $s'$ increases from 32 to 1536. 
The reduction rate for each method and each value of $s'$ is computed as
\begin{equation}
\text{Reduction Rate} = \frac{\text{Latency}_{\mathrm{Vanilla}} - \text{Latency}_{\mathrm{Method}}}{\text{Latency}_{\mathrm{Vanilla}}} \times 100\% ,
\end{equation}
where $\text{Latency}_{\mathrm{Vanilla}}$ denotes the latency of vanilla GAT at the same $s'$ value.
According to Table~\ref{tab:throughput_latency}, at smaller scales ($s' \leq 256$), all methods report sub 0.1~ms latencies, but GraphGHHA still achieves the lowest values, consistently remaining below 0.03~ms. At moderate scales, such as $s'=512$, GraphGHHA achieves 0.0502~ms latency, corresponding to a 75.8\% reduction compared to vanilla GAT. At the largest scale tested ($s'=1536$), sparsified GAT reduces latency to 0.316~ms (a 76.9\% reduction), while GraphGHHA reaches 0.146~ms, representing an 89.3\% reduction. These results highlight that GraphGHHA maintains consistently lower latency across all scales, with advantages that become more obvious as $s'$ increases.

\begin{figure}[ht]
  \centering
  \includegraphics[width=2.8in]{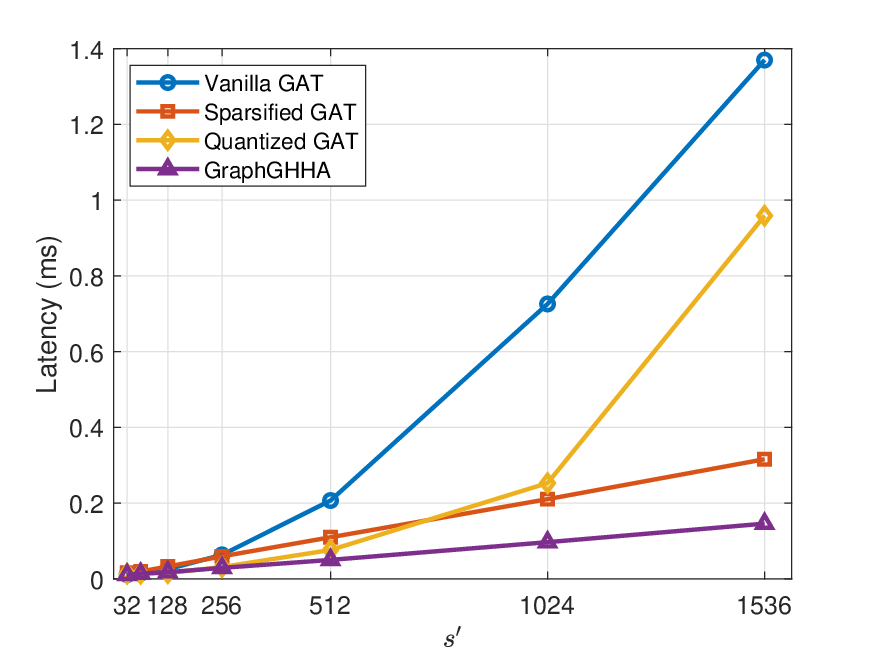}
  \caption{Average inference latency (ms) versus $s'$ for the four attention methods on the FCU graph.}
  \label{fig:latency_vs_s_hist}
\end{figure}

Fig.~\ref{fig:latency_vs_s_hist} shows the latency curves as a function of the node size $s^{\prime}$. The measured vanilla GAT latency increases from 0.0113~ms at $s^{\prime}=32$ to 1.37~ms at $s^{\prime}=1536$. Sparsified GAT has lower latency at larger scales, but is slower than vanilla GAT at the smallest sizes. Quantized GAT reaches 0.0764~ms at $s^{\prime}=512$ and 0.959~ms at $s^{\prime}=1536$. GraphGHHA exhibits approximately linear latency growth over the larger tested graph sizes and has the lowest reported latency at six of the seven evaluated sizes.
Table~\ref{tab:throughput_latency} and Fig.~\ref{fig:latency_vs_s_hist} show that GraphGHHA reduces inference latency relative to the evaluated methods' implementation, with larger relative gains on larger graphs.

\subsection{Detection Performance}

\begin{table*}[ht]
  \centering
  \caption{Detection precision (P), recall (R), and F1‐score vs.\ node size $s'$.}
  \label{tab:detection_vs_k}
  \small
  \begin{tabular}{c|ccc|ccc|ccc|ccc}
    \toprule
    & \multicolumn{3}{c|}{Vanilla GAT} 
    & \multicolumn{3}{c|}{Sparsified GAT} 
    & \multicolumn{3}{c|}{Quantized GAT} 
    & \multicolumn{3}{c}{GraphGHHA} \\
    $s'$ & P & R & F1 & P & R & F1 & P & R & F1 & P & R & F1 \\
    \midrule
    32 & 0.890 & 0.900 & 0.895 & 0.830 & 0.880 & 0.854 & 0.850 & 0.920 & 0.884 & 0.876 & 0.875 & 0.876 \\
    64 & 0.910 & 0.920 & 0.915 & 0.860 & 0.900 & 0.880 & 0.880 & 0.940 & 0.909 & 0.890 & 0.912 & 0.901 \\
    128 & 0.925 & 0.940 & 0.932 & 0.880 & 0.915 & 0.897 & 0.914 & 0.940 & 0.927 & 0.918 & 0.928 & 0.923 \\
    256 & 0.945 & 0.960 & 0.952 & 0.900 & 0.940 & 0.919 & 0.947 & 0.952 & 0.950 & 0.945 & 0.958 & 0.951 \\
    512 & 0.950 & 0.976 & 0.963 & 0.913 & 0.960 & 0.936 & 0.954 & 0.967 & 0.961 & 0.965 & 0.969 & 0.967 \\
    1024 & 0.959 & 0.983 & 0.971 & 0.920 & 0.965 & 0.942 & 0.970 & 0.976 & 0.971 & 0.974 & 0.978 & 0.976 \\
    1536 & 0.979 & 0.989 & 0.984 & 0.915 & 0.978 & 0.946 & 0.975 & 0.983 & 0.979 & 0.980 & 0.984 & 0.982 \\ 
    \bottomrule
  \end{tabular}
\end{table*}

Table~\ref{tab:detection_vs_k} presents precision (P), recall (R), and F1-score (F1) for four attention methods as $s'$ varies on the graph. The reported scores generally increase with graph size. Vanilla GAT reaches an F1-score of 0.984 at $s^{\prime}=1536$, while sparsified GAT, which restricts the retained neighborhood, reaches 0.946. Quantized GAT keeps the baseline neighborhood structure and reaches an F1-score of 0.979 at the same graph size. At larger graph sizes, GraphGHHA approaches vanilla GAT's detection performance and achieves the highest reported F1-score at $s^{\prime}=512$, with a score of 0.967. At $s^{\prime}=1536$, GraphGHHA achieves an F1-score of 0.982, compared with 0.984 for vanilla GAT. Together with the latency measurements, these results indicate a favorable accuracy-latency trade-off at larger graph sizes.

\begin{figure}[t]
\centering
\includegraphics[width=2.8in]{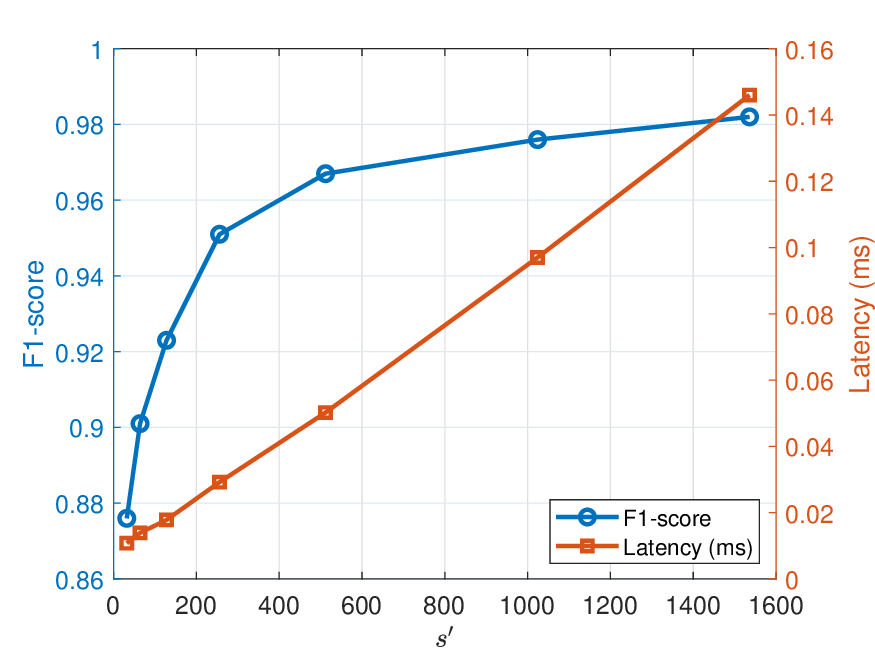}
\caption{GraphGHHA detection F1-score (left axis) and average latency (right axis) vs.\ $s'$.}
\label{fig:lsgat_tradeoff}
\end{figure}

To further investigate the relationship between $s'$, detection performance and inference latency for GraphGHHA, we plot F$1$‐score and average latency as functions of $s'$ in Fig.~\ref{fig:lsgat_tradeoff}. As $s'$ increases from 32 to 256, the F$1$-score increases steeply from 0.876 to 0.951, then more gradually to 0.982 at $s'=1536$. At the same time, the average latency increases nearly linearly. From 0.011 ms at $s'=32$ to 0.029 ms at $s'=256$, and to 0.146 ms at $s'=1536$. 

The analysis of the initial accuracy gains and latency growth with $s'$ suggests that $s'$ in the range $256\le s'\le512$ offers the best balance, delivering F$1$‐score above 0.95 with inference times under 0.05 ms. In our ablation studies, we fix $s'=1024$, since at this point GraphGHHA achieves an F$1$‐score of 0.976 (within 0.006 of its maximum) while requiring only 0.097 ms. This configuration provides near‐peak detection quality with minimal additional latency cost.

\begin{table}[ht]
  \centering
  \caption{Average per‐scenario F$1$‐score (\%) for GraphGHHA ($K=8$, $s'=1024$). The overall mean is 97.6\%.}
  \label{tab:per_scenario_f1}
  \small
  \begin{tabular}{l c}
    \toprule
    Scenario & Avg.\ F$1$‐score (\%) \\
    \midrule
    Damper Attacks (Leak \& Stuck) & 97.8 \\
    Valve Actuator Attacks (Leak \& Stuck) & 97.7 \\
    Sensor Bias ($\pm 2^\circ$C, $\pm 4^\circ$C) & 97.6 \\
    Coil Fouling (Air \& Water) & 97.5 \\
    Airflow Obstructions (Filter, Inlet, Outlet) & 97.8 \\
    Controller Attacks (Reverse \& Unstable) & 97.4 \\
    \midrule
    \textbf{Overall Average} & \textbf{97.6} \\
    \bottomrule
  \end{tabular}
\end{table}

\subsection{Per‐Scenario Detection Quality}
To further examine GraphGHHA’s robustness across different attack types, we group the evaluated FCU cases into six scenario families and compute the average F$1$‐score for each.  Table~\ref{tab:per_scenario_f1} presents these results in percent.

Damper attacks (leak and stuck) and airflow obstruction attacks each achieve the highest F$1$-score at 97.8\%, while controller attacks are slightly lower at 97.4\%.  Valve actuator attacks, sensor bias, and coil fouling register 97.7\%, 97.6\%, and 97.5\%, respectively.  The overall average F$1$‐score across all six scenario groups is 97.6\%, demonstrating that GraphGHHA maintains consistently high detection quality across the evaluated scenario families.

\subsection{Comparison with Alternative Graph Models and Unit Variants}
\label{sec:backbone_ablation}

\begin{table}[ht]
    \centering
    \caption{Accuracy and inference latency of graph
    models and unit variants at $s'=256$.}
    \label{tab:backbone_comparison}
    \label{tab:branch_comparison}
    \begin{tabular}{lcc}
        \toprule
        Method & Accuracy (\%) & Latency (ms) \\
        \midrule
        Vanilla GNN & 93.30 & 0.0146 \\
        GCN & 92.84 & \textbf{0.0137} \\
        Vanilla GAT & 95.20 & 0.0636 \\
        \midrule
        LAU-only & \textbf{97.01} & 0.0413 \\
        GAU-only & 92.39 & 0.0145 \\
        GraphGHHA & 95.02 & 0.0292 \\
        \bottomrule
    \end{tabular}
\end{table}

We compare GraphGHHA with vanilla GNN, graph convolutional network (GCN), and vanilla GAT to motivate graph attention acceleration, and with LAU-only and GAU-only models to examine the contributions of the two units to the overall performance of GraphGHHA. All models use $s'=256$, three processing blocks, and a hidden width of 64. The embedding, residual connections, feed-forward networks, pooling, and classifier share the same design.

Table~\ref{tab:backbone_comparison} reports the accuracy and inference latency of vanilla GNN, GCN, vanilla GAT, GraphGHHA, and its LAU-only and GAU-only models at $s'=256$. It shows that vanilla GAT achieves higher accuracy than vanilla GNN and GCN but has greater latency. GraphGHHA reduces GAT's latency by 54.9\%, with an accuracy decrease of 0.18 \%. These results motivate accelerating graph attention while retaining near-GAT detection performance.

According to Table~\ref{tab:backbone_comparison}, the LAU-only model achieves higher accuracy than GAU-only, whereas GAU-only has lower latency. GraphGHHA reduces latency by 30.5\% relative to LAU-only, with an accuracy decrease of 1.99 \%. Compared with GAU-only, it improves accuracy by 2.63 \% but incurs higher latency. Combining the LAU and GAU therefore provides a desirable accuracy-latency trade-off.

\subsection{Ablation Study}

\subsubsection{Head‐Count Ablation}
\begin{figure}[ht]
  \centering
  \includegraphics[width=2.8in]{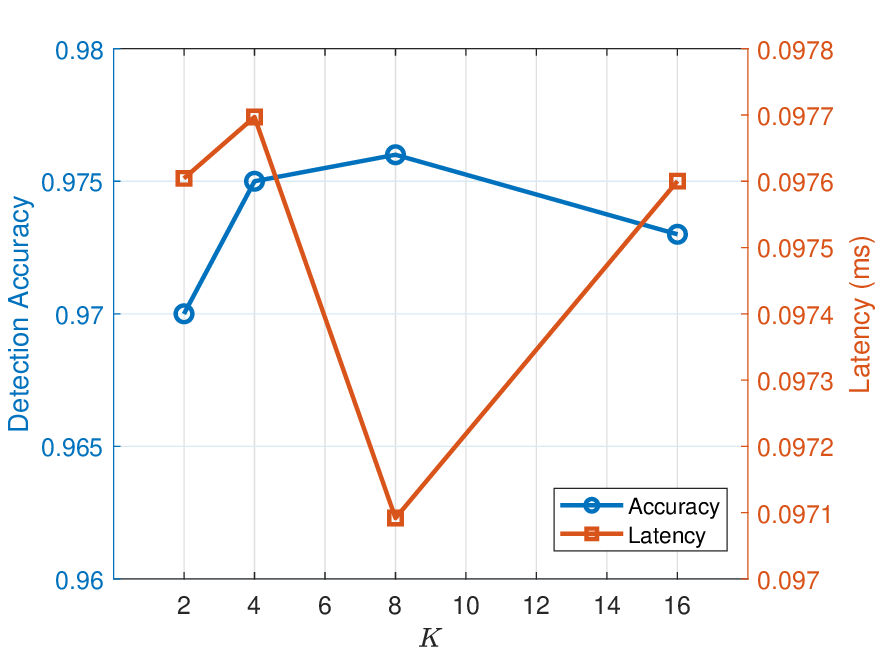}
  \caption{Head‐count ablation for GraphGHHA ($s'=1024$): detection F1-score vs.\ number of heads (left axis) and latency (right axis).}
  \label{fig:heads_ablation}
\end{figure}

To understand the effect of the number of attention heads $K$ on both detection quality and computational cost, we fix the neighbor-budget parameter to $n=8$ and vary $K\in\{2,4,8,16\}$.  All experiments use the GraphGHHA framework with $s'=1024$.  
Fig.~\ref{fig:heads_ablation} shows detection F1-score (left vertical axis) and average latency in milliseconds (right axis) as functions of $K$.  We observe that F1-score rises from 0.970 at $K=2$ to a peak of 0.976 at $K=8$, then slightly decreases to 0.973 at $K=16$. Increasing the head count beyond eight does not improve the reported F1-score. For the change of latency, it remains nearly constant around 0.097 ms across all head counts, showing little measured latency variation over the tested head counts.
These results suggest that $K=8$ provides the best trade-off, achieving the highest detection accuracy without significant impact on inference latency.  Accordingly, we select $K=8$ for all experiments.

\begin{table}[ht]
  \centering
  \caption{Impact of Numerical Precision ($s'=1024$, $K=8$). Latency reduction is measured relative to FP32.}
  \label{tab:ablate_precision}
  \small
  \begin{tabular}{lccccc}
    \toprule
    & Precision & Recall & F$1$ & Latency (ms) & Reduction(\%) \\
    \midrule
    FP32 & 0.976 & 0.980 & 0.978 & 0.152 & 0.0 \\
    FP16 & 0.974 & 0.978 & 0.976 & 0.097 & 36.2 \\
    BF16 & 0.970 & 0.975 & 0.973 & 0.098 & 35.5 \\
    \bottomrule
  \end{tabular}
\end{table}

\subsubsection{Effect of Numerical Precision}\label{quantization}
We compare full‐precision GraphGHHA (FP32), half‐precision GraphGHHA (FP16), and GraphGHHA (BF16) under fixed $s'=1024$, $K=8$. 
Compared with BF16, FP16 provides more significand precision. However, it has a smaller exponent range. In contrast, BF16 reduces the mantissa precision but retains the same exponent range as FP32. This design gives BF16 a much wider dynamic range.

The ablation study in Table \ref{tab:ablate_precision} demonstrates that simply switching GraphGHHA’s arithmetic from full 32-bit precision to 16-bit half precision reduces runtime with almost no loss in detection quality.  In FP32, each inference requires on average 0.152 ms and achieves an F$1$‐score of 0.978. Moving to FP16 cuts the latency to 0.097 ms, a 36.2\% reduction relative to FP32, while the F$1$-score only decreases by 0.002. This confirms that GraphGHHA can be further accelerated using low precision with a small decrease in the reported F1-score.

Adopting BF16 instead of FP32 achieves nearly the same latency (0.098 ms) as FP16, corresponding to a 35.5\% reduction, but incurs an additional slight drop in F$1$-score (0.978 to 0.973) due to its reduced mantissa width. In practice, choosing FP16 provides the best balance, which delivers over 1.5$\times$ acceleration compared to FP32 with a 0.2 \% decrease in F$1$-score, and the preferred format depends on hardware support and numerical stability.

\section{Conclusion}
In this paper, we proposed GraphGHHA, a graph-native hybrid-head acceleration layer for real-time attack detection in networked control systems. GraphGHHA is designed as a drop-in replacement for graph attention layers, combining an LAU for adjacency-constrained sparse attention and a GAU for system-wide mixing. 

We integrated GraphGHHA into an HVAC simulation and evaluated its detection performance under the considered abnormal operating scenarios. At $s^{\prime}=1536$, GraphGHHA reduces the reported inference latency by 89.3\% relative to the evaluated vanilla GAT baseline, while achieving an F1-score of 0.982 compared with 0.984. The configuration with $K=8$ and $s^{\prime}=1024$ achieves a reported F1-score above 0.97, demonstrating low model-execution latency with high detection performance in the evaluated setting.
By achieving near-baseline GAT detection performance in the evaluated HVAC setting, GraphGHHA bridges the gap between graph-based anomaly detection and practical deployment in cyber–physical control systems. Besides the HVAC system, the proposed architecture could be investigated in other automation domains, such as water distribution and power grid monitoring, where low latency and reliable anomaly detection are needed.

Future work will focus on extending GraphGHHA to dynamic graphs, and exploring hardware acceleration and self-supervised learning to enhance scalability and robustness.


\bibliographystyle{IEEEtran}

\bibliography{references.bib}



\vspace{11pt}

\vspace{11pt}

\vfill

\end{document}